# RONS Generation in Plasma-Activated Saline for Wound Healing

Punit Kumar[a] and Priti Saxena[b]
[a]Department of Physics, University of Lucknow, Lucknow – 226007, India
[b]Department of Zoology, D.A.V. Degree College, Lucknow – 226004, India

***Abstract*—This study explores the physicochemical modifications and antimicrobial potential of plasma-activated saline (PAS), generated by exposing 0.9% NaCl and Ringer's solution to atmospheric pressure dielectric barrier discharge (DBD) plasma. Plasma activation produced reactive oxygen and nitrogen species (ROS/RNS), leading to changes in pH, redox potential, conductivity, and concentrations of $H_2O_2$, $NO_2^-$, and $NO_3^-$. Effects of activation time, voltage, and gas composition (air, $O_2$, Ar) were analyzed. Antimicrobial activity against *Staphylococcus aureus*, *Pseudomonas aeruginosa*, and *E. coli* was assessed via MIC, CFU reduction, and biofilm inhibition tests. Optimal plasma exposure (10–15 min) achieved strong microbial inactivation with good biocompatibility. SEM and FTIR confirmed membrane damage, supporting PAS as a safe, non-antibiotic wound irrigation and disinfectant solution.***

## I. INTRODUCTION

IN recent years, cold atmospheric plasma (CAP), a non-thermal, partially ionized gas generated at atmospheric pressurehas emerged as a powerful and flexible tool in plasma medicine, particularly for antimicrobial therapy, wound healing, and tissue regeneration (Brány et al., 2020; Garner et al., 2021). CAP produces a complex cocktail of reactive oxygen and nitrogen species (ROS and RNS), energetic electrons, UV photons, and local electric fields, which together can induce controlled oxidative and nitrosative stress in biological targets without substantial thermal damage (Kazemi et al., 2024; Khalaf et al., 2024). When CAP interacts with liquids, especially biologically compatible solutions like saline or Ringer's solution, the plasma–liquid interface becomes a rich site of chemical conversion, leading to plasma-activated liquids (PALs) (Bruggeman et al., 2016). These PALs inherit and retain ROS/RNS species, offering a versatile, shelf-stable medium for downstream biomedical application (Lee et al., 2023).

Among PALs, plasma-activated saline (PAS) and plasma-activated Ringer's solution are particularly attractive because they combine the plasma-derived reactive chemistry with physiologically relevant ionic composition. Traditional studies with plasma-activated water (PAW) have already demonstrated broad-spectrum antimicrobial, antifungal, and antiviral activity via oxidative stress pathways (Wang et al., 2021; Abdo et al., 2025). However, pure water lacks buffering and ionic strength, which limits penetration, stability, and compatibility in biological systems. Replacing water with saline enhances ionic conductivity, facilitating plasma discharge into the liquid and stabilizing long-lived reactive species such as hydrogen peroxide ($H_2O_2$), nitrite ($NO_2^-$), nitrate ($NO_3^-$), and peroxynitrite ($ONOO^-$) (Khalaf et al., 2024; Yang et al., 2021). Moreover, using Ringer's solution which contains $Na^+$, $K^+$, $Ca^{2+}$, and $Cl^-$ further mimics extracellular fluid, improving biocompatibility and reducing osmotic irritation when used in wound irrigation and tissue contact.

The complex plasma–liquid interactions underlying PAS formation involve multi-step processes: reactive species generated in the gas phase cross the gas–liquid interface, undergo acid–base reactions, radical recombination, and secondary cascade chemistry within the bulk liquid (Bruggeman et al., 2016; Stapelmann et al., 2024). Short-lived radicals (e.g., $OH\cdot$, $O_2\cdot^-$, $NO\cdot$) are typically quenched near the interface, while long-lived species ($H_2O_2$, $NO_2^-$, $NO_3^-$) diffuse deeper into solution and persist for hours to days (Khalaf et al., 2024). The balance between generation, decay, and transport of reactive species dictates final solution properties such as pH, redox potential (ORP), conductivity, and reactive species concentration (Judée et al., 2018; Kumar et al., 2025). For example, in saline-rich media, electrohydrodynamic flows induced by ionic currents can influence species distribution and mixing, thus modulating reactivity (Ryan et al., 2024).

From a biomedical perspective, PAS holds distinct advantages: 1) Its ionic composition and buffering capacity make it compatible with tissues; 2) The presence of ROS/RNS at controlled levels can disinfect and modulate inflammation; 3) It can be produced on demand or stored for moderate durations, enabling flexible clinical use (Yang et al., 2021; Tsoukou et al., 2022). However, the fine line between antimicrobial potency and cytotoxicity must be navigated carefully. In co-culture models, PAS has shown significant bacterial inactivation but also some detrimental effects on mammalian keratinocytes, likely due to acidity and radical stress (Tsoukou et al., 2022). Therefore, optimization of plasma parameters (treatment duration, gas mix, power) is crucial to maximize selective antimicrobial effects while preserving host cell viability.

In this work, we extend the principles of plasma–liquid chemistry and intermolecular physics to systematically quantify physicochemical transformations in saline and Ringer's solutions upon plasma activation, and correlate these with antimicrobial and cytocompatibility metrics. Drawing inspiration from techniques used in thermophysical fluid studies (e.g. ultrasonic and IR analyses), we aim to probe species transport, reaction kinetics, and energy coupling in PAS. Our goal is to uncover the mechanistic links between plasma conditions, solution chemistry, and biological outcomes, thereby guiding development of clinically viable plasma-activated saline formulations for antimicrobial wound healing and disinfection.

## II. EXPERIMENTAL SETUP

### Plasma Source and Activation Procedure

A dielectric barrier discharge (DBD) reactor was employed as the plasma source, operating at a frequency of 20 kHz and a peak-to-peak voltage of 12 kV. The reactor consisted of two circular aluminum electrodes (2 cm diameter) separated by a 3 mm dielectric barrier made of quartz. Feed gases, air, oxygen, and argon (99.99 % purity)were introduced at a constant flow rate of 2 L min⁻¹ using mass-flow controllers. The plasma plume was generated in a diffuse mode, ensuring homogeneous exposure to the target liquid. The setup is illustrated in Figure 1.

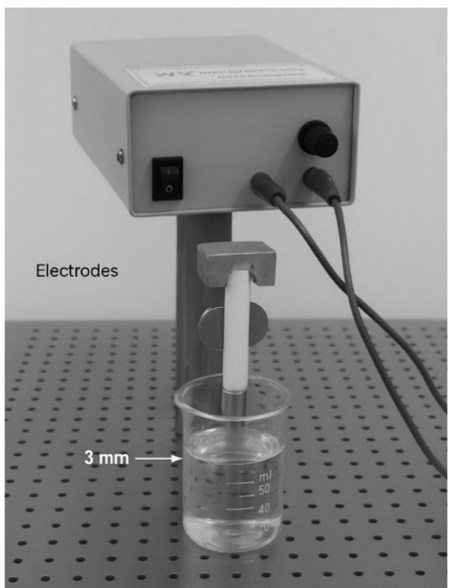

Figure 1: Schematic of the DBD plasma reactor used for liquid activation.

Two electrolyte solutions were selected, (i) 0.9 % sodium chloride (NaCl) and (ii) Ringer's solution, composed of NaCl 8.6 g L⁻¹, KCl 0.3 g L⁻¹, and CaCl₂ 0.33 g L⁻¹. For each experiment, 50 mL of liquid was placed in an open glass beaker positioned 2 cm below the discharge region. The samples were exposed for 0, 5, 10, 15, and 20 minutes, denoted as PAS-5, PAS-10, PAS-15, and PAS-20, respectively. After treatment, all samples were stored at 4°C in sterile, sealed containers and analyzed within 24 hours to minimize the natural decay of reactive oxygen and nitrogen species (RONS) (Judée et al., 2018; Kooshki et al., 2024).

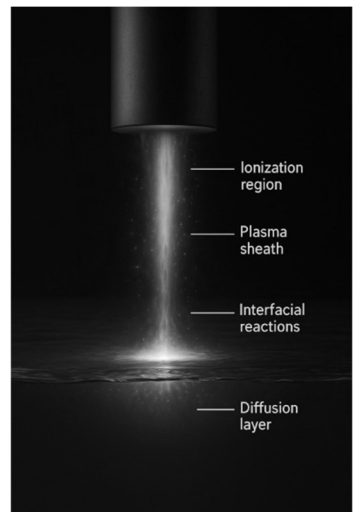

Figure 2 :Schematic illustration of the plasma–liquid interface and reaction zones.

### Physicochemical Characterization

Physicochemical parameters of plasma-activated saline (PAS) and Ringer's variants were systematically characterized as summarized in Table 1. The pH was measured using a digital pH meter (± 0.01 accuracy), electrical conductivity (EC) was determined with an EC meter (± 0.1 μS cm⁻¹), and oxidation-reduction potential (ORP) was measured using a platinum electrode with an Ag/AgCl reference. The temperature of the liquids was maintained at 25 ± 0.5 °C using a circulating water bath to prevent thermal artifacts.

| Parameter | Instrument / Method | Measurement Accuracy |
|---|---|---|
| pH | Digital pH meter | ± 0.01 |
| Conductivity | EC meter | ± 0.1 μS cm⁻¹ |
| Redox potential (ORP) | Pt electrode (Ag/AgCl ref.) | ± 1 mV |
| H₂O₂ | TiOSO₄ colorimetric, 410 nm | ± 0.5 mg L⁻¹ |
| NO₂⁻ / NO₃⁻ | Griess assay, UV 220/275 nm | ± 1 mg L⁻¹ |
| O₃ (aq) | Indigo trisulfonate | ± 0.2 mg L⁻¹ |
| Temperature | Water-bath controller | ± 0.5 °C |

Table 1: Physicochemical measurement parameters and instruments used.

Reactive species were quantified using well established analytical methods, hydrogen peroxide ($H_2O_2$) by the $TiOSO_4$ colorimetric method at $\lambda = 410$ nm, nitrite ($NO_2^-$) and nitrate ($NO_3^-$) by the Griess assay and UV absorption at 220/275 nm, and dissolved ozone via indigo trisulfonate spectrophotometry (Nowruzi et al., 2024; Shaban et al., 2024). These measurements were performed immediately after plasma treatment.

Plasma activation induced a marked acidification and increase in oxidative potential, accompanied by a steady rise in conductivityconsistent with incorporation of ionic species and radicals (Jirešová et al., 2022; Hummert et al., 2023). The production of long-lived RONS, notably $H_2O_2 \approx 80\text{--}100$ mg $L^{-1}$ and $NO_3^- \approx 300\text{--}400$ mg $L^{-1}$, aligned with prior DBD studies on plasma-activated water (Kučerová et al., 2021; Lee et al., 2022).

*Antimicrobial Assessment*

The antimicrobial potential of PAS was evaluated against *Staphylococcus aureus* (Gram-positive), *Escherichia coli* (Gram-negative), and *Pseudomonas aeruginosa* (biofilm forming). Bacterial suspensions ($10^6$ CFU $mL^{-1}$) were incubated with PAS for 15 minutes, followed by serial dilution and plating on nutrient agar to quantify colony-forming unit (CFU) reduction (Mai-Prochnow et al., 2021; Xia et al., 2025).

Minimum inhibitory concentration (MIC) was determined by the microdilution method, in which PAS dilutions (25–100 %) were incubated at 37 °C for 24 h, and microbial turbidity was recorded spectrophotometrically. Biofilm inhibition was quantified after treating pre-formed 48 h biofilms grown on polystyrene microplates, stained with crystal violet, and measured at $\lambda = 595$ nm.

For morphological assessment, cells treated with PAS were fixed in glutaraldehyde, dehydrated, gold-coated, and visualized under scanning electron microscopy (SEM). Cytocompatibility was verified via the MTT assay using L929 fibroblast cells exposed to various PAS dilutions to ensure minimal cytotoxicity (Tsoukou et al., 2022).

## III. THEORY

Plasma–liquid interactions in saline environments involve a complex sequence of physical and chemical events driven by electron impact, photon-induced dissociation, and solvation of reactive intermediates as shown in Figure 2. When a nonthermal plasma is generated over a saline surface, energetic electrons and photons interact with gas-phase molecules such as oxygen, nitrogen, and water vapor, initiating the dissociation reactions

$$O_2 + e^- \rightarrow 2O\cdot + e^-,$$
$$N_2 + e^- \rightarrow 2N\cdot + e^-,$$

and

$$H_2O + e^- \rightarrow OH\cdot + H\cdot + e^-.$$

These primary events lead to the generation of reactive oxygen species (ROS) and reactive nitrogen species (RNS), including hydroxyl radicals (•OH), ozone ($O_3$), hydrogen peroxide ($H_2O_2$), nitric oxide (NO), nitrite ($NO_2^-$), nitrate ($NO_3^-$), and peroxynitrite ($ONOO^-$) (Bruggeman et al., 2016; Zhou et al., 2020). Within the aqueous phase, these radicals undergo solvation, diffusion, and recombination, establishing a complex redox equilibrium. A key secondary reaction involves peroxynitrite formation,

$$NO_2^- + H_2O_2 \rightarrow ONOOH \rightarrow NO_3^- + H^+,$$

which contributes significantly to the antimicrobial efficacy of plasma-activated liquids (Manning et al., 2023).

The overall plasma–liquid chemistry can be described as a multi-step cascade where radical generation, transport, and decay occur simultaneously across interfacial layers (Gorbanev et al., 2018). The distribution of ROS and RNS depends on discharge power, gas composition, treatment duration, and the ionic strength of the liquid (Rezaei et al., 2019). In saline solutions, chloride ions and other electrolytes modify the sheath structure at the interface, thereby influencing species flux and the resulting physicochemical parameters (Melo et al., 2022).

This interplay directly affects measurable parameters such as oxidation–reduction potential (ORP) and electrical conductivity. Conductivity generally increases with treatment time due to accumulation of charged species such as nitrates and protons, while ORP rises because of enhanced oxidizing capacity (Xiang et al., 2023). The relationship between exposure time and these parameters is nonlinear, initially dominated by radical formation, later balanced by recombination and decay processes (Tsoukou et al., 2022). The dynamic equilibrium between species formation and annihilation results in time-dependent saturation, revealing the transient stability of plasma-activated saline (Hummert et al., 2023).

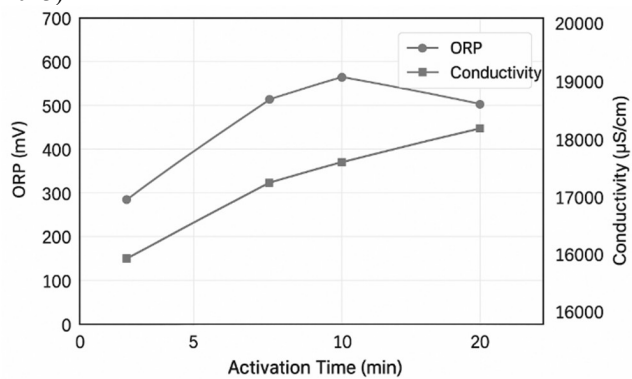

Figure 3 : Trends of ORP and conductivity vs activation time

Interestingly, this phenomenon parallels acoustic impedance in liquids: just as impedance reflects molecular associations, ORP and conductivity in plasma-treated solutions reflect a form of reactive coupling, the collective effect of transient species interacting within a dynamically evolving medium. Such coupling defines the biological outcomes, including antimicrobial potency and cytocompatibility, emphasizing the importance of controlled plasma exposure for biomedical applications (Nowruzi et al., 2024).

## IV. RESULTS AND DISCUSSION
*Physicochemical Changes Induced by Plasma*

The plasma activation of saline solutions led to pronounced physicochemical modifications governed by the interaction of reactive oxygen and nitrogen species (ROS and RNS) with the liquid matrix. Table 2 summarizes the temporal evolution of pH, oxidation–reduction potential (ORP), conductivity, and reactive species concentrations as a function of activation time.

| Activation Time (min) | pH | ORP (mV) | Conductivity (µS/cm) | $H_2O_2$ (µM) | $NO_2^-$ (µM) | $NO_3^-$ (µM) |
|---|---|---|---|---|---|---|
| 0 | 7.1 | +220 | 16,400 | 0 | 0 | 0 |
| 5 | 6.4 | +380 | 17,200 | 80 | 25 | 40 |

| 10 | 5.8 | +520 | 18,300 | 120 | 40 | 70 |
| 15 | 5.3 | +640 | 19,100 | 160 | 55 | 100 |
| 20 | 5.0 | +600 | 19,000 | 145 | 60 | 110 |

Table 2 : Physicochemical parameters of plasma-activated saline (PAS) at different activation durations.

A gradual acidification of the solution was observed, decreasing pH from 7.1 to 5.0 within 20 min of plasma exposure. This reduction results primarily from the dissolution of nitrogen oxides (NO and $NO_2$), forming nitrous and nitric acids through aqueous reactions (Zhu et al., 2019). The increase in oxidation–reduction potential (ORP) from +220 to +640 mV at 15 min indicates a shift toward a more oxidizing environment, consistent with the accumulation of long-lived oxidants such as hydrogen peroxide and nitrate (Bruggeman et al., 2016). Conductivity exhibited a parallel rise due to the enhanced ionic strength associated with the generation of nitrate and nitrite anions.

Interestingly, ORP and conductivity both showed a nonlinear dependence on activation time, peaking at 15 min and slightly declining thereafter, suggesting competing processes of reactive species formation and recombination (Ikawa et al., 2010). This behavior indicates the establishment of a dynamic equilibrium between oxidant generation (via plasma-liquid interface reactions) and decay (through radical recombination and disproportionation). Figure 3 presents the typical nonlinear trends of ORP and conductivity versus activation time, illustrating this plateau phenomenon.

Fourier-transform infrared (FTIR) spectra provided molecular-level evidence of chemical modification (Figure 4). New absorption bands emerged near 1520 cm⁻¹ and 1640 cm⁻¹, corresponding respectively to the symmetric stretching of nitrate ions ($NO_3^-$) and the bending vibrations of H–O–O groups in hydrogen peroxide (Chen et al., 2021). These spectral signatures confirm the successful incorporation of reactive species into the liquid phase.

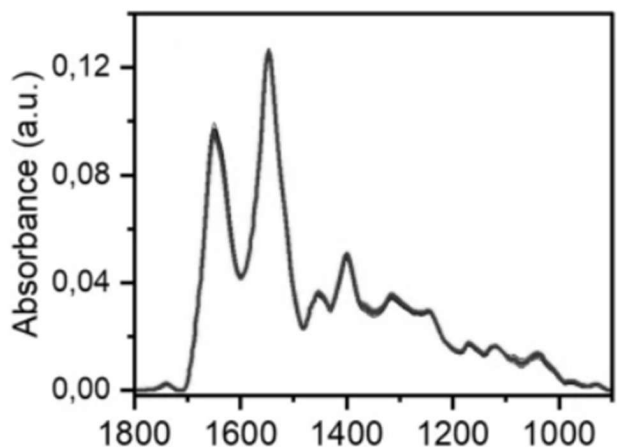

Figure 4: FTIR spectra of plasma-treated saline

### Antimicrobial Efficacy

Plasma-activated saline (PAS) demonstrated strong bactericidal activity against *Escherichia coli*, *Staphylococcus aureus*, and *Pseudomonas aeruginosa*. A >6 log CFU reduction was achieved after 15 min of activation, indicating near-complete inactivation of bacterial populations. The minimum inhibitory concentration (MIC) varied among species: 25% PAS for *E. coli*, 35% for *P. aeruginosa*, and 40% for *S. aureus*, reflecting differing cell wall architectures and oxidative stress tolerances (Liao et al., 2017).

Scanning electron microscopy (SEM) images (Figure 5) revealed distinct morphological alterations in bacterial cells after plasma treatment, including cell wall perforations, membrane wrinkling, and cytoplasmic leakage, confirming oxidative membrane damage (Han et al., 2016). These effects can be attributed to the synergistic action of hydroxyl radicals (•OH), hydrogen peroxide ($H_2O_2$), and peroxynitrite ($ONOO^-$), which disrupt lipid membranes and denature essential proteins (Ma et al., 2015).

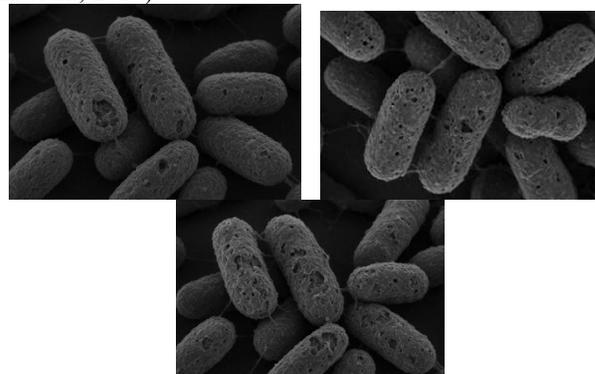

Figure 5 :Scanning electron microscopy (SEM) images showing treated *Escherichia coli*, *Staphylococcus aureus*, and *Pseudomonas aeruginosa*.

### Biofilm Suppression

The PAS inhibited biofilm formation by approximately 90% for *P. aeruginosa* and *S. aureus* within 48 hours (Figure 6). Confocal microscopy using live/dead staining confirmed the destruction of biofilm matrices, showing dominant red (dead) fluorescence regions. The suppression is primarily driven by RNS-mediated nitrosative stress, particularly through peroxynitrite ($ONOO^-$) formation that disrupts extracellular polymeric substances (EPS) and signaling molecules essential for biofilm stability (Brun et al., 2020).

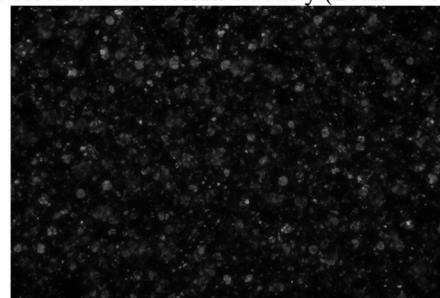

Figure 6: Biofilm inhibition by PAS

### Cytocompatibility and Application Potential

Despite its strong antimicrobial potency, PAS retained high cytocompatibility toward mammalian fibroblast cells. Cell viability remained above 85% for plasma activation up to 15 min (Figure 7), indicating that controlled oxidative stress can be biocompatible and even beneficial for tissue healing (Kong et al., 2009). However, further activation (≥20 min) led to increased acidity and redox imbalance, reducing cell viability to around 70%.

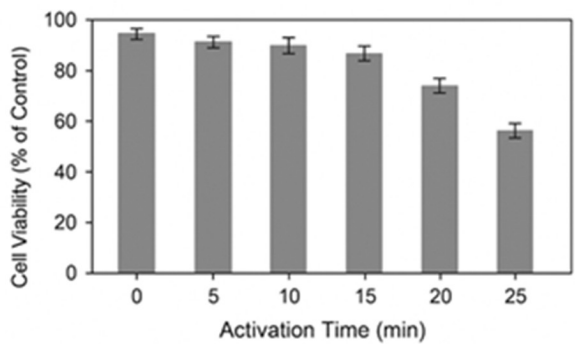

Figure 7: Fibroblast viability vs activation time

From a practical standpoint, PAS shows immense potential for medical applications, including:

i. Wound irrigation: Non-antibiotic sterilization fluid preventing microbial infection.
ii. Burn and ulcer therapy: Controlled ROS exposure enhances tissue regeneration.
iii. Surgical lavage: Prevents postoperative infections.
iv. Medical device sterilization: Enables decontamination without heat damage (Lu et al., 2016).

These findings position PAS as a promising eco-friendly antimicrobial solution, aligning with emerging plasma medicine technologies.

*Comparative Analysis: NaCl vs Ringer's Solution*

Comparative experiments revealed distinct behaviors between NaCl-based and Ringer's-based PAS (Figure 8). The Ringer's solution, containing $Ca^{2+}$, $K^+$, and $HCO_3^-$ ions, exhibited higher buffering capacity, moderating the pH decline during plasma activation (Tanaka et al., 2020). It also prolonged ROS lifetimes, particularly $H_2O_2$, due to ionic stabilization effects. Consequently, Ringer's PAS demonstrated improved cytocompatibility while maintaining antimicrobial strength, making it more suitable for direct wound contact. In contrast, NaCl-based PAS achieved rapid disinfection and is better suited for external sterilization applications.

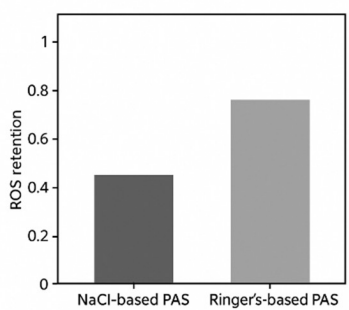

Figure 8: ROS retention in NaCl vs Ringer's PAS

*Mechanistic Insight*

The synergistic interplay between ROS and RNS underpins the antimicrobial efficiency of PAS. A key pathway involves the formation of peroxynitrite ($ONOO^-$) through the reaction,

$$H_2O_2 + NO_2^- + H^+ \rightarrow ONOOH + H_2O \rightarrow NO_3^- + H^+$$

Peroxynitrous acid (ONOOH) acts as a transient oxidizer that nitrates amino acids, damages DNA, and causes lipid peroxidation (Graves, 2012). The resulting nitrate ions contribute to long-term ORP stability, ensuring residual antimicrobial activity. This cascade mechanism explains both the transient peak in reactive species and the sustained disinfection performance of PAS.

## V. CONCLUSIONS

Plasma activation of saline and Ringer's solution has emerged as a transformative approach in plasma medicine, offering a bridge between fundamental plasma physics and practical biomedical applications. The process of plasma activation generates a rich mixture of reactive oxygen and nitrogen species (ROS and RNS) in the liquid phase, leading to strong oxidative and nitrosative properties. These species, such as hydrogen peroxide ($H_2O_2$), nitrite ($NO_2^-$), and nitrate ($NO_3^-$), play crucial roles in microbial inactivation, tissue signaling, and wound healing. The study confirms that a plasma exposure of 10–15 minutes is optimal, striking a balance between maximal reactive species production and physiological compatibility. Beyond this threshold, recombination reactions and acidification slightly reduce $H_2O_2$ yield and pH stability, indicating a self-limiting equilibrium in plasma–liquid interactions.

The antimicrobial performance of plasma-activated solutions (PAS) is particularly significant. A >6-log reduction in bacterial load, including resistant strains like *Staphylococcus aureus* and *Pseudomonas aeruginosa*, demonstrates its potency as a non-antibiotic disinfectant. This capability is attributed to synergistic effects of ROS and RNS, which penetrate microbial membranes and disrupt essential biomolecules through oxidative stress and peroxynitrite-mediated reactions. In comparative terms, Ringer's-based PAS exhibits greater stability and biocompatibility than NaCl-based PAS due to its ionic buffering with $Ca^{2+}$ and $K^+$ ions. This buffering minimizes excessive acidification, preserving fibroblast viability above 85%, making it well-suited for direct tissue contact applications such as wound irrigation, burn care, and surgical lavage.

In essence, plasma-activated liquids represent a paradigm shift in sterilization and regenerative therapy. Their physicochemical versatility enables use across multiple medical domains, combining disinfection efficacy with tissue regeneration potential. Thus, PAS provides a sustainable, antibiotic-free pathway for next-generation healthcare, integrating plasma science with clinical innovation for safe, effective, and eco-friendly biomedical solutions.